\documentclass[graybox]{svmult}

\usepackage{type1cm}        
\usepackage{makeidx}         
\usepackage{graphicx}        
\usepackage{multicol}        
\usepackage[bottom]{footmisc}

\usepackage{newtxtext}       %
\usepackage{newtxmath}       

\makeindex             

\usepackage{float}
\usepackage{cite}
\usepackage{enumitem}
\newlist{steps}{enumerate}{1}
\setlist[steps, 1]{label = Step \arabic*:}

\usepackage{comment}
\usepackage{hyperref}

\begin{document}

\title*{Correlations among Game of Thieves and other centrality measures in complex networks}
\titlerunning{Our experience with Game of Thieves} 
\author{Annamaria Ficara, Giacomo Fiumara, Pasquale De Meo, and Antonio Liotta}
\authorrunning{A. Ficara, G. Fiumara, P. De Meo, and A. Liotta} 
\institute{Annamaria Ficara \at University of Palermo, Palermo, Italy, \email{aficara@unime.it}
\and Giacomo Fiumara \at University of Messina, Messina, Italy, \email{gfiumara@unime.it}
\and Pasquale De Meo \at University of Messina, Messina, Italy, \email{pdemeo@unime.it}
\and Antonio Liotta \at Free University of Bozen-Bolzano, Bolzano, Italy, \email{antonio.liotta@unibz.it}}

%
%
\maketitle

\abstract{Social Network Analysis (SNA) is used to study the exchange of resources among individuals, groups, or organizations. The role of individuals or connections in a network is described by a set of centrality metrics which represent one of the most important results of SNA. Degree, closeness, betweenness and clustering coefficient are the most used centrality measures. Their use is, however, severely hampered by their computation cost. This issue can be overcome by an algorithm called Game of Thieves (GoT). Thanks to this new algorithm, we can compute the importance of all elements in a network ({\em i.e.} vertices and edges), compared to the total number of vertices. This calculation is done not in a quadratic time, as when we use the classical methods, but in polylogarithmic time. Starting from this we present our results on the correlation existing between GoT and the most widely used centrality measures. From our experiments emerge that a strong correlation exists, which makes GoT eligible as a centrality measure for large scale complex networks.}

\section{Introduction} 
\label{sec:1}
SNA studies groups \cite{Fortino2020} of individuals and it can find an application in a lot of areas such us organizational studies, psychology, economics, information science and criminology \cite{Ficara2020, CALDERONI2020113666}. Social Networks (SNs) like Facebook and Twitter have grown exponentially providing new challenges for the application of SNA methods. 

The definition of the so-called centrality measures represents one of the most important results of SNA. These set of measures describe the role of single individuals (or single connections) with respect to their network of relationships and can be used to identify the most influential people. These people have the potential of controlling the information flow inside a network and, for this reason, they have a great practical relevance. 
Thanks to the use of the main centrality metrics, such us degree, closeness, betweenness, and clustering coefficient, we can increase our understanding of a network.

When we compute centrality measures on on-line SNs, we are facing the problem about the big size of the data. This problem can be overcome by using a new algorithm called Game of Thieves (GoT)  \cite{Mocanu}. GoT computes the centrality of both vertices and edges in a network with respect to the total number of vertices. This computation is done in polylogarithmic time, while the classical centrality measures need at least a quadratic time. 

GoT owes its name to the protagonists of the game who are a multitude of thieves whose main purpose is to steal diamonds. 

The basic idea is to make an overlap between a heterogeneous system like a complex network \cite{Pace2019} and a homogeneous artificial system which has two key elements: a group of thieves and a set of vdiamonds ({\em i.e.} virtual diamonds). At the beginning, each vertex is artificially endowed with vdiamonds and wandering thieves. If a thief does not carry any vdiamond, his state is ``empty''. If he carries a vdiamond, his state becomes ``loaded''. If the thief state is ``empty'', he wanders in search of vdiamonds. The thief picks randomly a neighbor of the vertex in which it is located, he moves to this new vertex and, if he finds a vdiamond, he fetches it. Then, he follows back the same path used in search of vdiamonds and brings the vdiamond back to his home vertex. At this point the vdiamond becomes available for the other thieves who can steal it.
At the beginning of the game, there is the same number of thieves and vdiamonds in each vertex. Then, GoT proceeds in epochs. At each epoch, all thieves move from their current location to the next one. When they find or deposit a new vdiamond, their state (``loaded'' or ``empty'') changes.

Encouraged from the superior performance of GoT respect to the state-of-art algorithms, we decided to investigate whether GoT can be used to compute vertices and edges centrality.
This amounts to investigate whether and to which extent exists a correlation between GoT and some classical centrality measures.

Correlation is a bivariate analysis through which we can study the association between two variables. This kind of analysis takes into account the strength of this relationship between pairs of variables and its direction. The value of the correlation coefficient can vary from $-1$ to $+1$. A perfect degree of association (positive or negative) between two variables is indicated by a value of $\pm 1$. The relationship between pairs of variables becomes weaker when the value of the correlation coefficient goes towards $0$. 
The most used types of correlations are Pearson correlation, Spearman and Kendall rank correlations.

We have done a lot of experiments computing these three correlation metrics on different types of networks both artificial and real. We used three classes of simulated networks: {\it Erd\"os Rényi} (ER) random graphs, {\it small-world} (SW) and {\it scale-free} (SF) networks. For each class, we randomly generated different networks which have $1,000$ to $15,000$ vertices and $4,970$ to $1,125,545$ edges. Then, we have taken into account three networks from real-world: Dolphins ($62$ vertices, $159$ edges, unweighted), High Energy ($8,361$ vertices, $15,751$ edges, weighted, disconnected) and Internet ($22,963$ vertices, $48,436$ edges, unweighted).
Our experiments show that there is a strong negative correlation among GoT and the main centrality metrics like degree, betweenness and closeness; while there is no correlation between GoT and the clustering coefficient with the exception of the small-world networks in which we can find a strong positive correlation.

\section{Related Literature}  
\label{sec:2}

Centrality measures describe the position of an individual in a network in relation to the complete network and to the other individuals in the same network. Some centrality metrics identify the most influential and prestigious actors in a network \cite{Freeman, Bonacich2, Scott, Wasserman}, some others indicate the social influence of an individual with respect to others in a network \cite{Friedkin}, others evaluate the integration of each individual into a network \cite{Valente}. Most recently, a new algorithm called Game of Thieves has been developed \cite{Mocanu}. It is a method which is able to compute the importance of both vertices and edges, which are the elements of a network, and to complete this computation in polylogarithmic time with respect to the total number of vertices.

Degree centrality, closeness centrality, betweenness centrality and clustering coefficient can be considered as the most frequently used centrality metrics. The first three measures were proposed by Freeman \cite{Freeman}, whereas the clustering coefficient was defined by Watts and Strogatz \cite{Watts}. 
In our work, we try to answer to an often asked, but rarely answered, question that is: are these centrality measures correlated? If there exists a high correlation between the centrality metrics, we can expect they have a similar behavior in statistical analyses and for this reason the development of multiple measures seems to be redundant. If there is not high correlation, we can conclude that they are unique measures which can be associated with different outcomes. 
But, we are not only interested in the correlation between the most used centrality metrics. We want to answer to an other question: are these centrality measures correlated with Game of Thieves? If we find that they behave similarly, we can use GoT in the computation of individuals' centrality in very large networks, considerably reducing the execution time of this computation.

Many researchers carried out studies on the correlations between centrality measures. 

Bolland \cite{Bolland} made a correlation analysis on four centrality measures: degree, closeness, betweenness, and continuing flow. He considered three criteria that are robustness, face validity and sensitivity. He underlined the similarity between closeness, degree and continuing flow and a relative difference between this three indices and the betweenness centrality. The high intercorrelations among the first three indices produced a considerable redundancy for the used dataset which was increased with the introduction of random error into the data. Then, the author chose the continuing flow as the best model and a useful companion to the betweenness.

Rothenberg et al. \cite{Rothenberg} compared eight centrality measures analyzing people risky behaviors in an area of low prevalence for HIV transmission. These measures were: three forms of information centrality ({\em i.e.} measures of centrality which make use of all paths between pairs of points) \cite{STEPHENSON19891}, eccentricity, mean, and median ({\em i.e.}, three distance measures), and degree and betweenness centrality. Their studies showed an high correlation among these eight centrality measures. In particular, there was an high correlation among the three distance measures and the three information measures, but there was a weaker correlation among these measures and degree and betweenness. Degree and betweenness were highly correlated, but both were less correlated with the three forms of information centrality which were highly correlated among themselves.

Faust \cite{Faust} used a subset of the data from Galaskiewicz's study \cite{Galaskiewicz} regarding relationships between CEOs, clubs and boards and examined correlations among several centrality measures. He used centrality measures such us degree, eigenvector, closeness, betweenness to compute the centrality of an event, and flow betweenness used to identify central clubs. Then, he studied the correlation among these metrics founding correlation coefficients between $0.89$ and $0.99$. 

Valente and Forman \cite{Valente} discovered two new centrality measures know as integration and radiality. They examined correlations among these two measures, in-degree, out-degree, closeness, betweenness, flow and density. They used the ``Sampson Monastery'' and the ``Medical Innovations'' datasets. Their analysis showed that integration was correlated with in-degree and radiality was correlated with out-degree. A further study on the correlation revealed that these new metrics were similar but distinct from closeness, betweenness and flow.

In a more recent study, Valente et al. \cite{Valente2008} choose the most commonly used centrality measures such as degree, in-degree, out-degree, betweenness, s-betweenness, closeness-in, closeness-out, s-closeness, integration, radiality and eigenvector. They empirically investigated the correlation among them finding out that degree had the strongest overall correlations. Eigenvector centrality had the next highest average correlation. Similar correlations were founded among betweenness, symmetrized closeness, in-degree and out-degree. The lowest average correlation was discovered between directional closeness measures, in-closeness and out-closeness. 

Li et al. \cite{Li2015} first studied the Pearson correlation between centrality measures and the similarity ranking for vertices. Then, they introduced a new centrality measure known as the degree mass. They found that betweenness, closeness, and eigenvector were strongly correlated with the degree, the 1st-order degree mass and the 2nd-order degree mass, respectively, in both artificial and real networks. Then, they demonstrated that eigenvector and the 2nd-order degree mass had a larger Pearson correlation coefficient respect to eigenvector and a lower order degree mass. 

Ronqui and Travieso \cite{Ronqui2015} studied the correlation between pairs of centrality measures in two artificial networks and several real networks. Their analysis showed that these metrics were usually correlated. A stronger correlation could be found in the artificial networks with respect to real networks. Moreover, the strength of the correlation between the centrality measures varied from network to network. For this reason, they proposed a centrality correlation profile as a way to characterize networks. This profile consisted of the values of the correlation coefficients between the centrality metrics of interest.

Grando et al. \cite{Grando2016} showed through their experiments that vertex centrality measures such as information, eigenvector, subgraph, walk betweenness and betweenness could identify vertices in all kinds of networks with a performance at 95\%. Considerably lower results could be achieved using other metrics. In addition, they demonstrated that several pairs of centrality metrics evaluate the vertices in a very similar way ({\em i.e.} their correlation coefficient values were above $0.7$). 

Shao et al. \cite{Shao2018} uses degree to approximate closeness, betweenness, and eigenvector. They first demonstrated that rank correlation performed better than the Pearson one in scale-free networks. Then, they studied the correlation between centrality metrics in real networks. At the end, they demonstrated that largest betweenness and closeness vertices could be approximated by the largest degree vertices. This approximation was not valid for the largest eigenvector vertices.

Oldham et al. \cite{Oldham2019} used $212$ different real networks and calculated correlations between $17$ different centrality measures. The relationship between these correlations and the variations in network density and global topology was examined together with the possibility for vertices to be clustered into distinct classes according to their centrality profiles. Their analysis showed that there was a positive correlation among the centrality measures. The strength of these correlations could vary across networks, and network modularity played a key role in driving these cross-network variations. 


\section{Background}
\label{sec:3}

\subsection{Centrality Measures} 
\label{subsec:1}
Centrality is a core concept for the SNA. A SN is a set of people interconnected by social ties, {\it e.g.}, friendship or family relationships \cite{Sakr}. 
It can be represented using a graph $G = (V, E)$ where $V$ is a set of vertices (also called nodes, actors) and $E \subseteq V \times V$ is a set of edges (also called links, ties). A graph is called {\it undirected} when all the edges are bidirectional, {\it directed}  when the edges have a specific direction. Given a directed edge $e = (u, v) \in E$, we can say that $v$ is the head of $e$, $u$ is the tail and $v$ is adjacent to $u$. Specific graph types can be used depending on the specific SN. We can represent a SN like Facebook with an undirected graph because in this case friendship relationships are reciprocal. Instead, we can use directed graphs to describe SNs like Twitter which use following relationships and require the use of edges with a specific direction.

A SN can be also defined as a {\em weighted graph} $G = (V, E, W)$ where $V$ is the set of vertices, $E \subseteq V \times V$ is the set of edges, and $W: E \rightarrow R_{++}$  is a set of positive weights defined on each edge.

{\it Degree Centrality} (DC) \cite{Freeman} is used to evaluate the local importance of a vertex and it is one of the simplest centrality measures; given a vertex $u$ the degree centrality $DC(u)$ of $u$ is as follows:
 $$ DC(u)  = \sum\limits_{w=1}^{v} a_{uw} $$
where $v$ is the number of vertices in $G$, $a_{uw} = 1$ if and only if there exists $(u,w) \in E$, $0$ otherwise.

{\it Betweenness Centrality} (BC) \cite{Brandes} measures how important the role of a vertex is in the propagation of informations. Some vertices, in fact, act as bridges between different parts of a graph and for this reason they can block the flow of informations from one region to other. Specifically, the (shortest-path) betweenness $BC(u)$ of a vertex $v$ is the sum of the fraction of all-pairs shortest paths that pass through $u$ and it defined as follows:
$$ BC(u) =  \sum\limits_{x,y \in V} {\sigma (x,y | u) \over \sigma (x,y)} $$
where $ \sigma (x, y)$ is the number of shortest paths between an arbitrary pair of vertices $x$ and $y$, and $ \sigma (x, y | u)$ is the number of shortest paths which connect $x$ and $y$ by passing through the vertex $u$.

{\it Closeness Centrality} (CL) \cite{Freeman} measures the ``proximity'' between a vertex and all other vertices in a graph $G$. The closeness centrality of a vertex $u$ is the reciprocal of the sum of the shortest path distances from $u$ to all other vertices in $G$, normalized by $v-1$:
$$ CL(u) = {v - 1 \over  \sum\limits_{w=1}^{v-1} d(u, w)} $$

{\it Clustering Coefficient} (CC) measures how connected a vertex neighbors are to one another. For unweighted graphs, the clustering of a vertex $u$, denoted by $CC(u)$, is the fraction of possible triangles through that vertex that exist,

$$ CC(u) = \frac{2 T(u)}{D(u)(D(u)-1)}, $$
where $T(u)$ is the number of triangles through vertex $u$ and $D(u)$ is the degree of $u$.
$CC(u) = 1$ if every neighbor connected to a vertex $u$ is also connected to every other vertex within the neighborhood.  $CC(u) = 0$ if no vertex that is connected to $u$ connects to any other vertex that is connected to $u$.


\subsection{Game of Thieves} 
\label{subsec:2}

{\it Game of Thieves} \cite{Mocanu} is a new centrality measure to compute the centrality of vertices and edges in a graph $G = (V, E)$, where $V$ is the set of vertices, and $E$ is the set of edges. 

As mentioned in Sect.~\ref{sec:1},  the leading actors in the game are wandering thieves. If a thief carry a vdiamond, his state is ``empty''. If a thief does not carry a vdiamond, his state is ``loaded''.  

In order to understand how this measure works, we have to define some notation:

\begin{itemize}
\item $\Phi_0^v$ is the initial number of vdiamonds in vertex $v \in V$ at epoch $T=0$; 
\item $\Phi_T^v$ indicates the number of vdiamonds in vertex $v \in V$  at epoch $T$ ({\em i.e.} after GoT has run for $T$ epochs); 
\item $\Psi^e_T$ is the number of ``loaded'' thieves passing through an edge $e \in E$ at epoch $T$;
\item $\Gamma_v$ is the set of vertices connected by an edge with vertex $v, \forall v \in V$; 
\item $\Omega_{vu} \ge 0$ is the weight of the edge which connects the vertex $v \in V$ and $u \in V$;  
\item $Y_t$ is a dynamic list which contains the vertices visited by a thief $t$, useful to keep the path of $t$ in his search for vdiamonds.
\end{itemize}

If the state of a thief $t$ is ``empty'', the following operations will be sequentially performed in any epoch $ep$:

\begin{steps}
\item $a$ randomly picks a vertex $u \in \Gamma_v$, where $v$ is its actual location, with a probability $p_{vu} = {\Omega_{uv} \over \sum_{v \in \Gamma_v}\Omega_{vu}}$.
\item $t$ moves from his home vertex $v$ to vertex $u$. 
\item If $u \in Y_t$, then all the vertices after $u$ in $Y_t$ are removed from the list.
\item If $u \notin Y_t$, then $u$ is added to the end of $Y_t$.
\item If $\Phi_{ep}^u \textgreater 0$, then $t$ takes one vdiamond and changes his state to ``loaded''.
\item $\Phi_{ep}^u$ decreases by one vdiamond.
\end{steps}

If the state of a thief $t$ is ``loaded'', the following steps will be sequentially performed in any epoch $ep$:
\begin{steps}
\item $t$ moves from the last vertex $v$ from $Y_t$, which is his actual location, to the last but one vertex $u$ from $Y_t$.
\item $v$ is removed from $Y_t$.
\item $\Psi^e_{ep}$ increases by one, i.e, edge $e$ from $v$ to $u$ increases.
\item If $u$ is the home vertex of $t$, $t$ unloads the vdiamond, and sets his state to ``empty''.
\item $\Phi_{ep}^u$ increases by one vdiamond.
\end{steps}

The game runs for a duration of $T$ epochs. The number of epochs to stop the algorithm is conventionally $T = \log^3 \lvert V \rvert$. 

\begin{figure}[t]
\numberwithin{figure}{subsection}
\sidecaption
\includegraphics[width=1.03\linewidth]{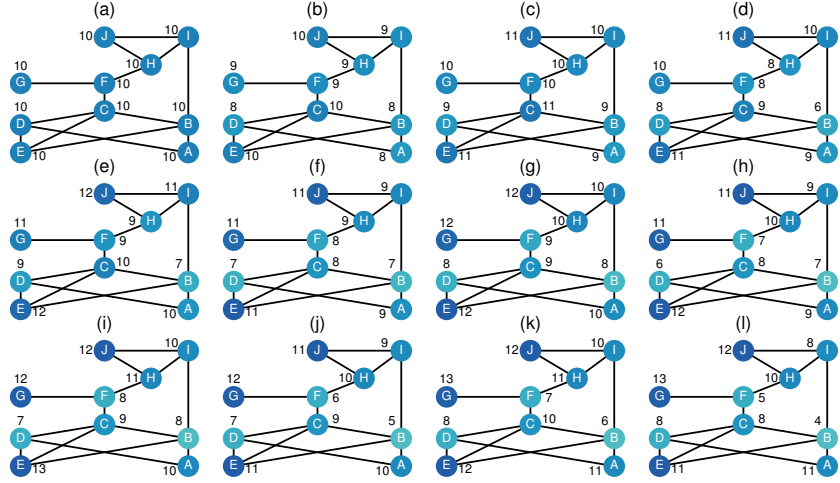}
\caption{\textbf{GoT in action.} GoT behavior over $T=12$ epochs on a simple unweighted network with $10$ vertices. The initial parameters are $\Phi_0^v=10$ and $1$ thief per vertex. The numbers on the side of each vertex show the number of vdiamonds, $\Phi_{ep}^v$ , in any vertex $v$ at epoch $ep$, where $ep = 1, 2, 3, 4, 5, 6, 7, 8, 9, 10, 11, 12$ epochs in subplots \textbf{a,b,c,d,e,f,g, h, i, j, k, l} respectively.}
\label{fig:GoT}       
\end{figure}

Figure \ref{fig:GoT} shows snapshots of GoT in action on a simple network with $10$ vertices. We can observe the thieves' behavior and consequently the number of vdiamonds on each vertex $v$ after $T = \log^3 \lvert 10 \rvert \approx 12$ epochs.

When the game stops the centrality of each vertex $v$ is computed as: $$ \bar{\Phi}_T^v = \frac{1}{T}\sum_{ep=0}^{T}\Phi_{ep}^v$$
This measure also refers to the average number of vdiamonds present at a vertex $v$, after the game has run for a duration of $T$ epochs. An important vertex is indicated by a small $\Phi_T^v$ value, while a less important vertex is denoted by a high $\Phi_T^v$ value. This is because a lot of thieves visit the most central vertices which will are quickly depleted, while few thieves visit the less central vertices which will not be depleted.

Then, the centrality of each edge $e$ is also computed as: $$\bar{\Psi}^e_T =  \frac{1}{T}\sum_{ep=0}^{T}\Psi_ep^e$$
This measure also refers to the average number of thieves who carry a vdiamond ({\em i.e.} in ``loaded'' state) passing through an edge $e$ after $T$ epochs. The most important edges are indicated by a high $\Psi^e_T$ values, while the less important edges are denoted by lower $\Psi^e_T$ values.

The computational complexity of GoT, $O(GoT)$, is bounded by $O(log^2|V|) < O(GoT) < O(log^3|V|)$. 

It's easy to guess from the description of the game that each vertex in the network is independent from the others. An high level of parallelization can be achieved in a traditional parallel computing environment, such as MPI. We can also think about a graph partitioning algorithm in which each vertex or a group of vertices can do their own computations. GoT seems to be a fully distributed algorithm. Table~\ref{tab:1} shows how GoT represents a great step forward in terms of time complexity with respect to centrality algorithms such us degree, betweenness, closeness and clustering.

\begin{table}[!t]
\numberwithin{table}{subsection}
\caption{Comparison of five centrality algorithms using computational complexity.}
\label{tab:1}       

\begin{tabular}{p{5.7cm}p{5.7cm}}
\hline\noalign{\smallskip}
\textbf{Algorithm} & \textbf{Computational complexity}  \\
\noalign{\smallskip}\svhline\noalign{\smallskip}
Degree Centrality & $O(|V|)$ \\
Betweenness Centrality & $O(|V||E|)$ \\
Closeness Centrality & $O(|V^3|)$ \\
Clustering Coefficient & $O(V^2)$ \\
Game of Thieves & $O(log^2|V|) < O(GoT) < O(log^3|V|)$ \\
\noalign{\smallskip}\hline\noalign{\smallskip}
\end{tabular}
\end{table}

\subsection{Correlation Coefficients} 
\label{subsec:3}

The correlation coefficient is a statistical measure of the strength of the relationship between two variables. The values of the coefficient can vary from $-1.0$ to $1.0$. A number greater than $1.0$ or less than $-1.0$ implies an error in the correlation measurement. A correlation of $-1.0$ means that there is a perfect negative correlation, while a correlation of $1.0$ shows a perfect positive correlation. A correlation of $0.0$ indicates no relationship between the two variables.

{\it Pearson $r$ correlation coefficient} \cite{Chen} is the most used correlation metric and it measures the degree of association between two linearly related variables. It is computed according to the following formula:

$$ r = { {s \sum ab - \sum(a)(b)} \over { \sqrt{ [s \sum a^2 - \sum (a^2)] [s \sum b^2 - \sum (b^2) ] }} } $$

where $r$ is the Pearson correlation coefficient, $s$ is the number of observations, $\sum ab$ is the sum of the products of $a$ and  $b$ scores, $\sum a$ is the sum of $a$ scores, $\sum b$ is the sum of $b$ scores, $\sum a^2$ is the sum of squared $a$ scores and  $\sum b^2$ is the sum of squared $b$ scores.

{\it Spearman rank correlation coefficient} \cite{Spearman} is a non-parametric measure of rank correlation. It measures the degree of relationship between two variables. The only hypothesis required is that the two variables can be ordered and, if possible, continued. This coefficient is computed according to the following formula:

$$ \rho = {1 - {{6 \sum d_{i}^2 } \over {s (s^2 -1)}}} $$

where $\rho$ is the Spearman rank correlation, $d_i$ is the difference between the ranks of corresponding variables and $s$ is the number of observations.

{\it Kendall rank correlation coefficient} \cite{Kendall} is a non-parametric test used to measure the strength of association between two variables. If we consider two samples, $x$ and $y$, where each sample size is $s$, $s(s-1)/2$ will be the total number of pairings with $x y$.  
This coefficient is computed according to the following formula:

$$ \tau = { {s_c - s_d} \over { \frac{1}{2}s(s - 1)}}$$

where $s_c$ is number of concordant pairs and $s_d$ is number of discordant pairs.

\subsection{Complex Networks}  
\label{subsec:4}

\ \\
{\bf Random networks}. 
A random network may be described simply by a probability distribution, or by a random process which generates it. The {\it Erdos–Rényi model} is one of two closely related models to generate random networks. There are two variants of the Erd\"os R\'enyi model \cite{erdos59a}. The first chooses one of all possible networks $G(v, E)$ with $v$ vertices and $E$ edges, where each network has an equal probability. This could be done by choosing $V$ edges from the $\binom{v}{2}$ possible edges. Second variant $G(v, p)$  \cite{gilbert1959} starts with an initial set of $v$ unconnected vertices and includes edges with probability $p$. It can easily be deduced that each network with $v$ vertices and $E$ edges is equally likely with probability:
$$ p^E (1-p)^{\binom{v}{2}-E} $$ 

\begin{figure}[t]
\sidecaption[t]
\includegraphics[width=1.05\linewidth]{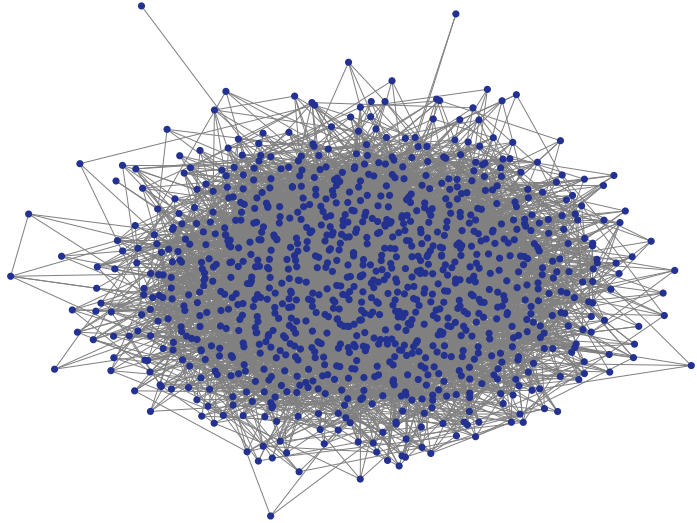}
\caption{\textbf{Random network}. ER model created using the variant $G(v, p)$ with $v = 1000$ and $p=0.01$ probability for edge creation.}
\label{fig:erdos}       
\end{figure}

In this paper, we used the second variant $G(v, p)$ of the ER model. In each experiment, we have chosen the number of vertices $v$ between $1,000$ (see Figure \ref{fig:erdos}) and $15,000$, and a probability for edge creation $p=0.01$.

\ \\
{\bf Small-world networks}. 
A small-world network \cite{Duan2017} is characterized by a high degree of local clustering (like regular lattices). It also possess short vertex-vertex distances. This network model was proposed by Watts and Strogatz \cite{Watts} and it interpolates between these two extremes by taking a regular lattice and randomly rewiring some of its edges. 

Newman and Watts \cite{NEWMAN1999341} proposed a variation of the Watts and Strogatz model.
Given a network defined as a graph $G(V,E)$, where $V$ is the set of vertices and $E$ is the set of edges, the {\it Newman-Watts-Strogatz small-world model} (NWS) is defined as follows:
\begin{steps}
\item {\it Ring Creation}. Creation of a ring over $v$ vertices in which each vertex $u \in V$ is connected with the $k$ closest neighbors. If $k$ is odd, $u$ is connected with the nearest $k-1$ neighbors.
\item {\it Edge rewiring}. For each edge $(u, w) \in E$, in the underlying $v$-ring with $k$ nearest neighbors, a new edge $(u, w)$ is added, with randomly-chosen existing vertex $w$ and probability $p$. 
\end{steps}
Compared with Watts-Strogatz model, the random rewiring increases the edges number because new edges are added and no edges are removed.

\begin{figure}[t]
\sidecaption[t]
\includegraphics[width=1.05\linewidth]{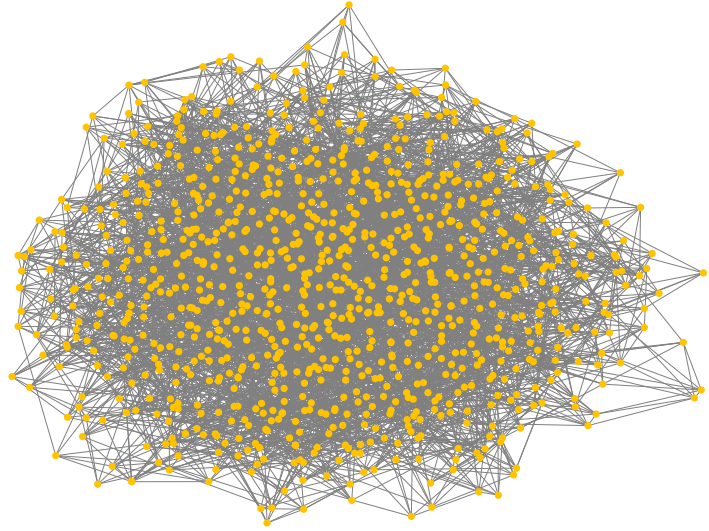}
\caption{\textbf{Small-world network}. NWS small-world network model with $v=1000$ vertices, each of which is joined with its $k=6$ nearest neighbors in the ring topology, and $p=0.6$ probability of rewiring each edge.}
\label{fig:small}       
\end{figure}

In this paper, we used the NWS model. In each experiment, we have chosen the number of vertices $v$ between $1,000$ (see Figure \ref{fig:small}) and $15,000$, $k=6$ neighbors with which connect each vertex $u$ in the ring topology, and a probability $p=0.6$ of rewiring each edge.

\ \\
{\bf Scale-free networks}.
A scale-free \cite{Duan2017} network is characterized by a degree distribution ({\em i.e.} the distribution of the number of vertices that have a particular degree) which decays like a power law \cite{Barabasi}. Given a network defined as a graph $G(V,E)$, where $V$ is the set of vertices and $E$ is the set of edges, the scale-free network model of {\it Barab\'asi and Albert} (BA) is defined as follows:
\begin{steps}
\item {\it Initial condition}. The network consists of $v_0$ vertices and $e_0$ edges. 
\item {\it Growth}. One vertex $u$ with $e$ edges is added at each step. Time $t$ is the number of steps. 
\item {\it Preferential attachment (PA)}. Each edge of $u$ is attached to an existing vertex $w$ with the following probability:
$$ P_i = \frac{D(w)}{\sum\limits_{u \in V}^{} D(u)} $$

The defined probability is proportional to the degree of vertex $u$.

Holme and Kim \cite{Holme2002} proposed a SF network model with two main characteristics: a perfect power-law degree distribution and a high clustering. To incorporate the second one, which is a peculiarity of the SW model, the authors modified the above BA algorithm by adding the following step: \\

\item {\it Triad formation (TF)}. If an edge $(u, w)$ was added in the PA step, an edge from $u$ to a neighbor of $w$ (chosen randomly) is added. If all neighbors of $w$ were already connected to $u$ ({\em i.e.} there are no pair to connect), a PA step is done instead. 
\end{steps}

\begin{figure}[t]
\sidecaption[t]
\includegraphics[width=1.05\linewidth]{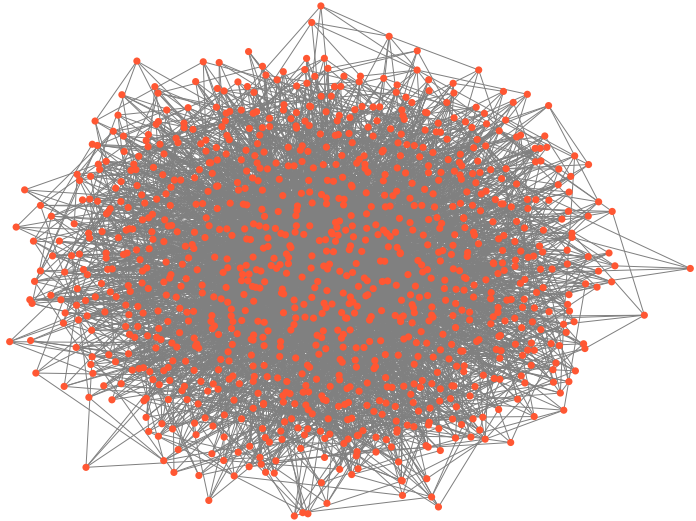}
\caption{\textbf{Scale-free network}.  Extended BA model by Holme and Kim with $v = 1000$ vertices, $e = 5$ random edges to add for each new vertex and $p = 0.3$ probability of adding a triangle after adding a new edge.}
\label{fig:scale}       
\end{figure}

In this paper, we used the BA model with the fourth extra step to generate scale-free networks. In each experiment, we have chosen the number of vertices $v$ between $1,000$ (see Figure \ref{fig:scale}) and $15,000$, we add $5$ random edges for each new vertex $u$, and we have chosen a probability $p = 0.3$ of adding a triangle after we have added each of these random edge.

\ \\

{\bf Real networks}. In this paper we used three real networks: Dolphins, High Energy and Internet. The corresponding datasets have been downloaded from Mark Newman’s website. 

\begin{figure}[t]
\sidecaption[t]
\includegraphics[width=1.05\linewidth]{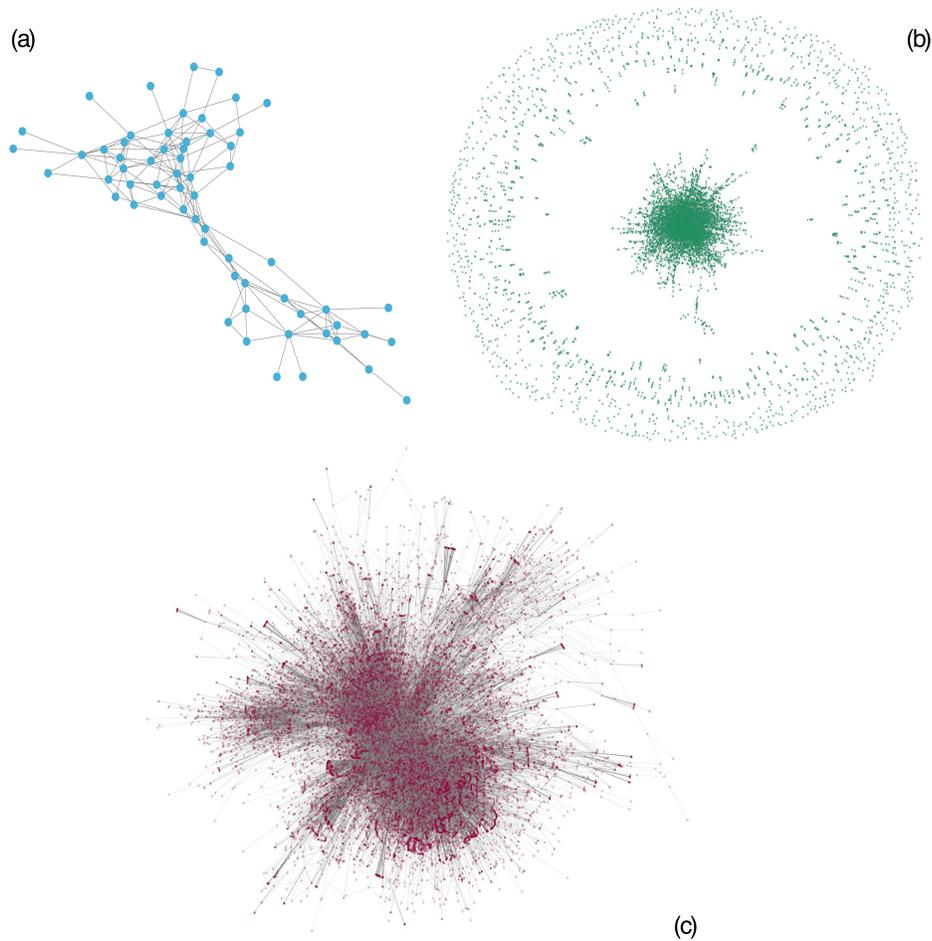}
\caption{\textbf{Real networks}. \textbf{(a)}: {\it Dolphins} social network with $62$ vertices ({\em i.e.} dolphins) and $159$ edges ({\em i.e.} frequent associations).  \textbf{(b)}: {\it High Energy} theory collaborations network with $8,361$ vertices ({\em i.e.} scientists) and $15,751$ edges ({\em i.e.} connections existing if the scientists have authored a paper together). \textbf{(c)}: {\it Internet} network with $22,963$ vertices ({\em i.e.} autonomous systems - AS) and $48,436$ edges ({\em i.e.} connections between AS).}
\label{fig:reals}       
\end{figure}

The {\it Dolphins} social network is an undirected and unweighted network of the relationships between the bottlenose dolphins (genus Tursiops) living in a community in New Zealand \cite{Lusseau}. The dolphins have been observed between $1994$ and $2001$. This network is composed of $62$ vertices which are the bottlenose dolphins and $159$ edges which are the frequent associations (see Figure \ref{fig:reals}-(a)).

The {\it High Energy} theory collaborations is an undirected and weighted network of co-authorships between scientists who posted preprints on the High-Energy Theory E-Print Archive between January $1$, $1995$ and December $31$, $1999$ \cite{Newman2}. This network is composed of $8,361$ vertices which are scientists and $15,751$ edges which are connections existing if the scientists have authored a paper together (see Figure \ref{fig:reals}-(b)).

The {\it Internet} network was created by Mark Newman from data for July $22$, $2006$ and is not previously published. It was reconstructed from BGP tables posted by the  University of Oregon Route Views Project. This network is a snapshot of the structure of the Internet at the level of autonomous systems (AS), {\em i.e.} collections of connected IP routing prefixes controlled by independent network operators.
It is an undirected and unweighted network in which the vertices are $22,963$ AS and the edges are $48,436$ connections between AS (see Figure \ref{fig:reals}-(c)).

\section{Correlation Analysis} 
\label{sec:4}

We investigated the correlations among the centrality measures introduced in Subsect.~\ref{subsec:1}, in both artificial and real-world networks described in Subsect.~\ref{subsec:4}. The network models include the SF networks, the SW networks and the ER random networks.
For each class, we randomly generated five unweighted networks. Each network had between $1,000$ and $15,000$ vertices.  Each SF network had between between $4,970$ and $74,959$ edges. Each SW network had between $4,810$ and $71,826$ edges. Each ER network had between between $5028$ and $1,125,545$ edges. 
The real-world networks include three networks from different domains: the {\it Dolphins} social network, the {\it High Energy} theory collaborations and the {\it Internet} network.

For the implementation of the centrality measures such as degree, closeness, betweenness, and clustering coefficient ,we used Python and NetworkX library \cite{SciPyProceedings11}. For GoT we used the implementation by D. C. Mocanu \cite{Mocanu} which is available on GitHub (\href{https://github.com/dcmocanu/centrality-metrics-complex-networks}{github.com/dcmocanu/centrality-metrics-complex-networks}), setting $1$ thief and $\Phi_0^v = \lvert V \rvert$ vdiamonds per vertex. We let GoT to run for $T=\log^3 \lvert V \rvert$ epochs. NetworkX was also used to generate the artificial networks and to perform our experiments with the real networks.

\begin{figure}[t]
\numberwithin{figure}{section}
\sidecaption
\includegraphics[width=1.02\linewidth]{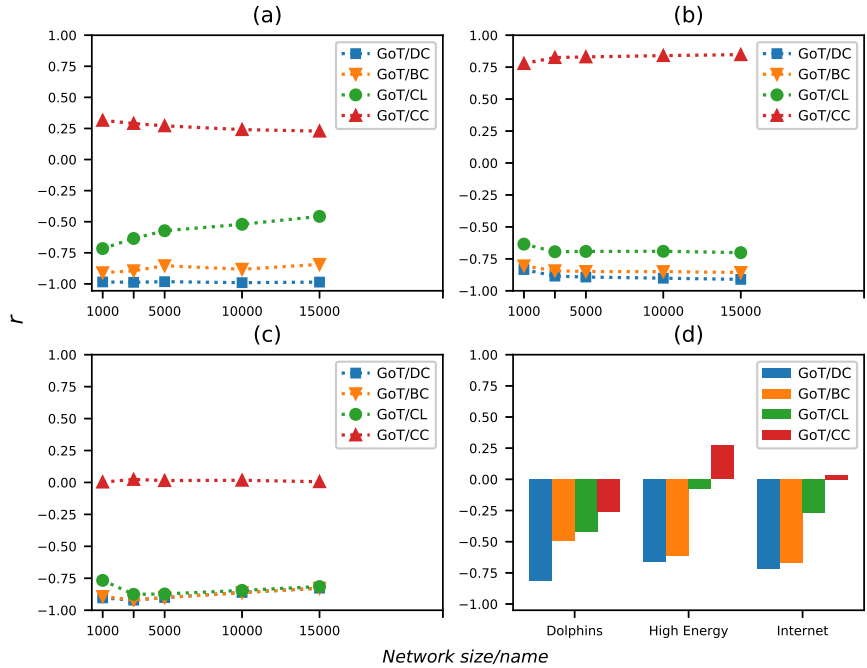}
\caption{\textbf{Pearson correlation coefficient $r$} between GoT and Degree (Blue), GoT and Betweenness (Orange), GoT and Closeness (Green), GoT and Clustering Coefficient (Red) as a function of the {\it Network size}, in SF networks \textbf{(a)}, SW networks (\textbf{b)}, ER random graph \textbf{(c)} and as a bar chart for real networks \textbf{(d)}. In the artificial networks, the size is between $1000$ and $15000$ vertices.} 
\label{fig:pearson}       
\end{figure}

\begin{figure}[t]
\sidecaption
\includegraphics[width=1.02\linewidth]{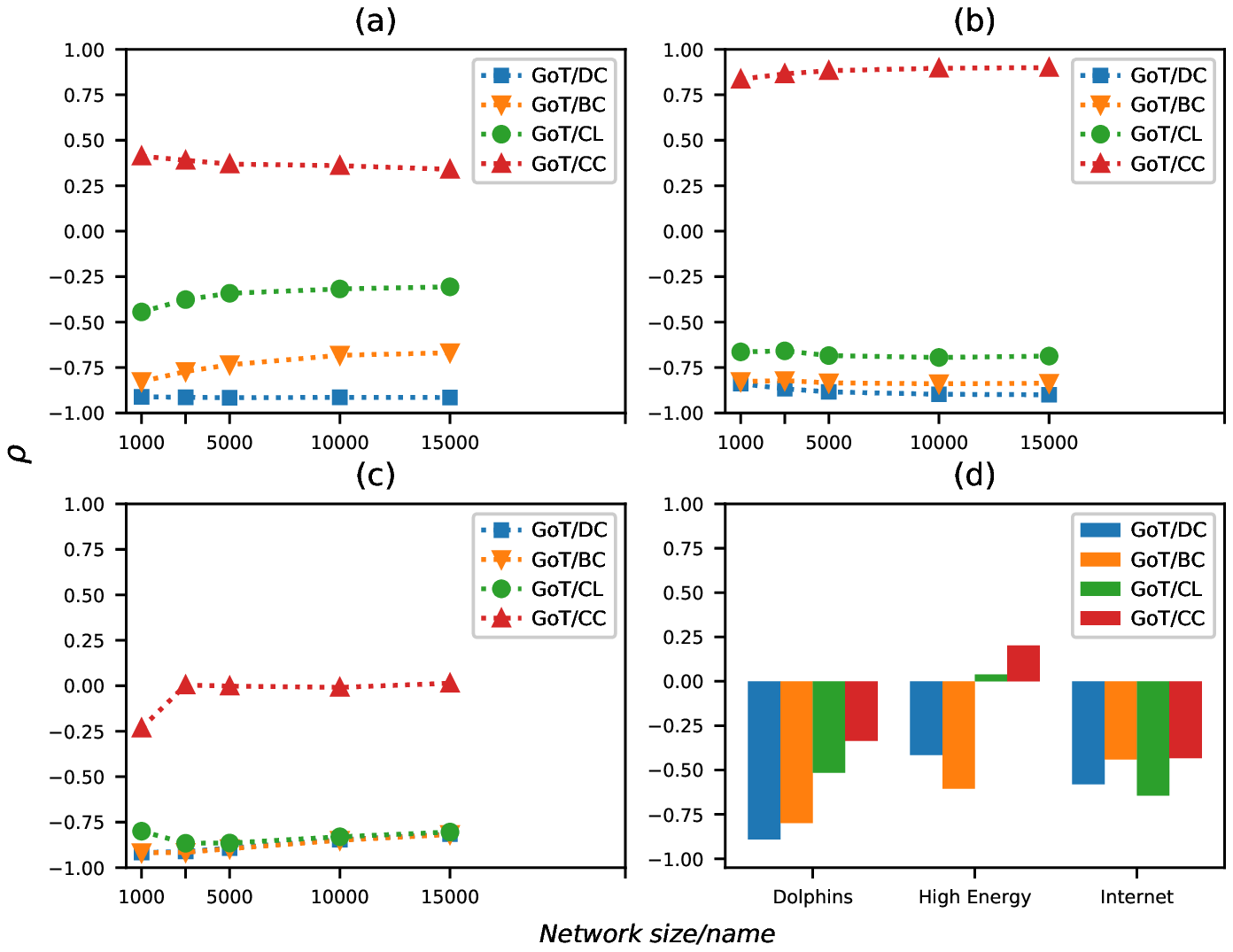}
\caption{\textbf{Spearman rank correlation coefficient $\rho$} between GoT and Degree (Blue), GoT and Betweenness (Orange), GoT and Closeness (Green), GoT and Clustering Coefficient (Red) as a function of the {\it Network size}, in SF networks \textbf{(a)}, SW networks \textbf{(b)}, ER random graph \textbf{(c)} and as a bar chart for real networks \textbf{(d)}. In the artificial networks, the size is between $1000$ and $15000$ vertices.} 
\label{fig:spearman}       
\end{figure}

\begin{figure}[t]
\sidecaption
\includegraphics[width=1.02\linewidth]{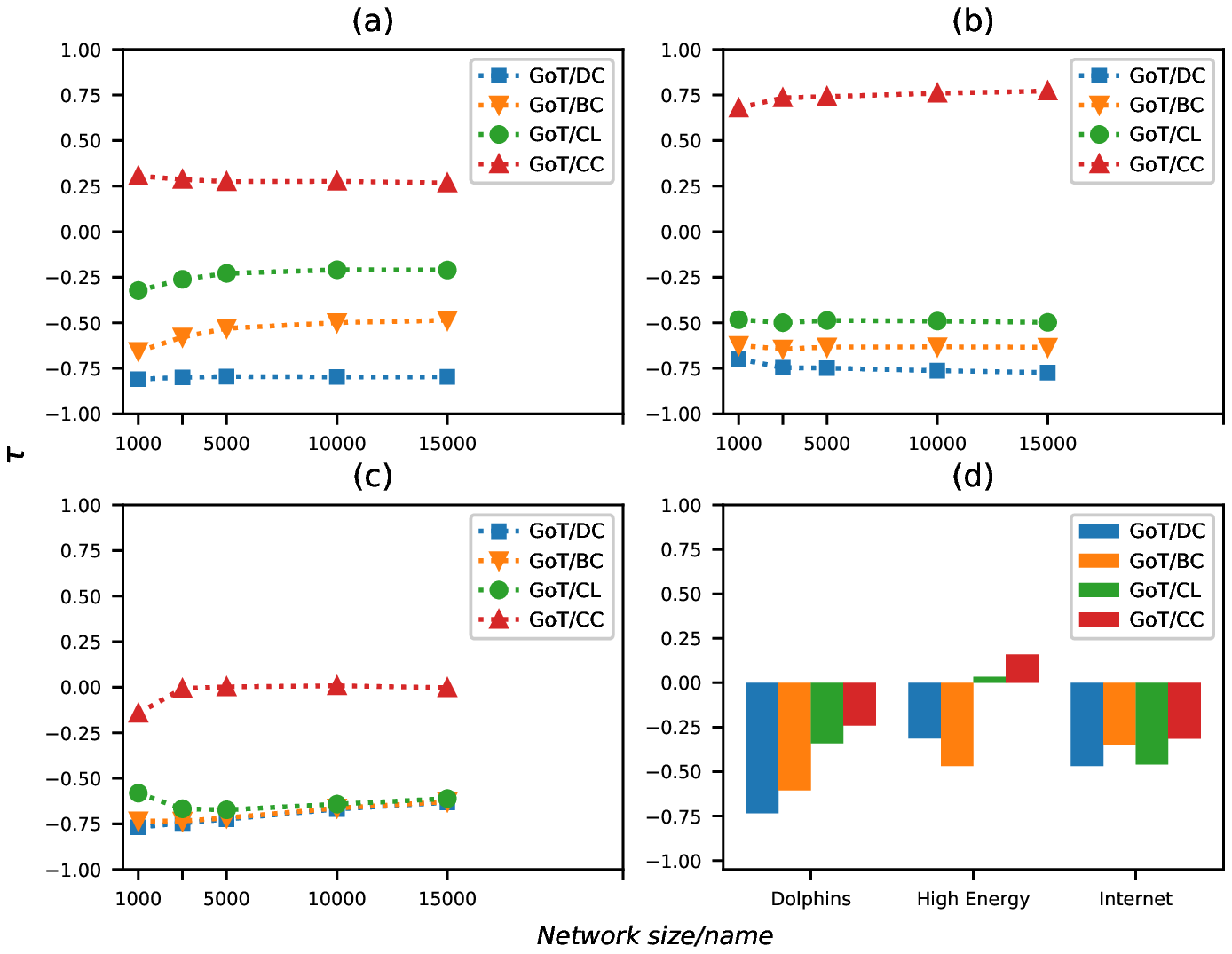}
\caption{\textbf{Kendall rank correlation coefficient $\tau$} between GoT and Degree (Blue), GoT and Betweenness (Orange), GoT and Closeness (Green), GoT and Clustering Coefficient (Red) as a function of the {\it Network size}, in SF networks \textbf{(a)}, SW networks \textbf{(b)}, ER random graph \textbf{(c)} and as a bar chart for real networks \textbf{(d)}. In the artificial networks, the size is between $1000$ and $15000$ vertices.} 
\label{fig:kendall}       
\end{figure}

The results of the {\it Pearson correlation coefficient} $r$ are presented in Figure \ref{fig:pearson}, the {\it Spearman rank correlation coefficient} $\rho$ in Figure \ref{fig:spearman} and the {\it Kendall Rank Correlation coefficient} $\tau$ in Figure \ref{fig:kendall}, with the growth of networks' sizes. Small deviations of rank correlation coefficients can be observed when the size of the networks is rather small. However, when networks grow big enough, the deviations are not visible anymore, especially for the rank correlation coefficients.
Spearman correlation coefficient $\rho$ was much higher than Pearson correlation coefficient $r$ and so more capable of capturing the underlying ranking correlation between GoT and the other measures.
Moreover, we can observe that $\rho$ is always larger than $\tau$, but there is no distribution difference between these two coefficients.

GoT and degree centrality have the strongest negative correlation.
GoT and betweenness centrality also exhibit a large negative correlation.
GoT and closeness centrality are negative correlated, but this correlation is less than that between GoT and both degree and betweenness.
GoT and clustering coefficient centrality have no correlation in most cases.
In ER networks, we can observe the strongest and almost identical negative correlation among GoT and degree, betweenness and closeness.
In SW networks, we can observe a very strong and unique positive correlation between GoT and the clustering coefficient.
Real networks are more complex than the artificial ones, but also in this case the correlation among GoT and degree centrality is confirmed to be the strongest one.

\section{Conclusions} 
\label{sec:5}

In this work we examined the correlation between well known and recently proposed centrality measures in real and artificial networks, {\em i.e.} scale-free, small-world and Erd\"os-Rényi networks. If two centrality measures have a strong correlation, it means there is the possibility of approximating the metric with the highest computational complexity using the other. We used the Pearson correlation coefficient, the Spearman and Kendall rank correlation coefficients to study the correlations between the centrality metrics. An important finding is that the degree and the betweenness are strongly correlated with the new metric Game of Thieves. Also the closeness centrality is correlated with GoT but it's some sort of weaker correlation. The clustering coefficient and the Game of Thieves have a strong positive correlation only in SW networks.

We have done a correlation analysis observing the correlation coefficients when the number of vertices in both artificial and real networks increases. As future work, we want to make an analysis on artificial networks taking into account the increase of the number of edges when the number of vertices does not change. 

Moreover, in this work, we focus on measures of vertex centrality and consequently on GoT's capability to compute the vertices centrality in a network. As future work, we want to apply the GoT algorithm to the case of edge centrality making a correlation analysis with the state-of-the-art measures of edge centrality. The centrality of an edge reflects its contribute spreading messages over a network, as short as possible, and we can use it as a tool for the community detection \cite{de2014mixing, MeoFFP13, MeoFFR12}.

We can conclude that the GoT algorithm represents a step forward compared to the classical centrality algorithms which have at least a quadratic computational complexity an it can be used instead of degree, betweenness and closeness centrality when we want to compute the centrality of a vertex in a very large network.

%
%
\bibliographystyle{spmpsci}
\bibliography{mybib}

\end{document}